\documentclass[slac_one]{revtex4}
\usepackage{graphicx}
\usepackage{fancyhdr}
\begin{document}

\title{{\small{Hadron Collider Physics Symposium (HCP2008),
Galena, Illinois, USA}}\\ 
\vspace{12pt}
B Physics Theory for Hadron Colliders} 

%

\author{G. Buchalla}
\affiliation{Ludwig-Maximilians-Universit\"at M\"unchen, 
Fakult\"at f\"ur Physik,\hspace*{8cm} \\
Arnold Sommerfeld Center for Theoretical Physics, 
D-80333 M\"unchen, Germany}
%

\begin{abstract}
A short overview of theoretical methods for $B$ physics at
hadron colliders is presented. The main emphasis is on the theory of
two-body hadronic $B$ decays, which provide a rich field of investigation
in particular for the Tevatron and the LHC. The subject holds both
interesting theoretical challenges as well as many opportunities
for flavor studies and new physics tests.
A brief review of the current status and recent developments is given.
A few additional topics in $B$ physics are also mentioned.
\end{abstract}

\maketitle

\thispagestyle{fancy}


\section{INTRODUCTION}\label{sec:intro}
Hadron colliders, such as the Tevatron at Fermilab and the upcoming
LHC at CERN, produce large amounts of $B$ hadrons of all varieties,
$B_u$, $B_d$,$B_s$ and $B_c$ mesons, and $b$-flavored baryons.
Their numerous decay channels probe the flavor sector of the Standard
Model (SM) and of any scenario conceived to go beyond it. 
Of special interest are loop-induced processes and CP-violating observables,
for example $B$-$\bar B$ mixing, $B\to\psi K_S$, $B\to\rho\rho$,
$B_s\to\phi\phi$, $B\to K^*\mu^+\mu^-$ or $B_s\to\mu^+\mu^-$,
which determine CKM angles or have sensitivity to new physics through
virtual particles (or both).

A major challenge for theory is to disentangle the quark-level flavor
physics from the QCD dynamics and to control the nonperturbative sector
of QCD as far as possible. Several tools are available for this task:
\begin{itemize}
\item
perturbative calculations of inclusive rare decays ($B\to X_s\gamma$)
based on the heavy-quark expansion.
\item
$SU(2)$, $SU(3)$ flavor symmetries based on the smallness of the
light quark masses $m_u$, $m_d$, $m_s\ll\Lambda_{QCD}$. 
\item
factorization of hadronic two-body and exclusive rare FCNC decays based
on the heavy-quark limit $m_b\gg\Lambda_{QCD}$.
\item
lattice-QCD computations, especially suited for ``static'' quantities
(decay constants $f_B$, bag factors $B_B$, $B\to\pi$ form factors at
small recoil) \cite{Kronfeld:2008zz}.
\item
QCD sum rules on the light cone (LCSR), in particular for nonperturbative
quantities with fast light hadrons (distribution amplitudes $\Phi_\pi(x)$,
$B\to\pi$ form factors at large recoil) \cite{Ball:2004ye,Ball:2004rg}.
\end{itemize}
These tools have been developped and refined over the years, a process
that is still continuing. The methods are complementary and interdependent.
For example, QCD sum rule or lattice results of hadronic form factors
and decay constants are needed as input for factorization calculations,
the latter in turn provide us with a framework to estimate $SU(3)$
breaking in amplitude relations based on flavor symmetries.

\begin{figure*}[t]
\centering
\includegraphics[width=135mm]{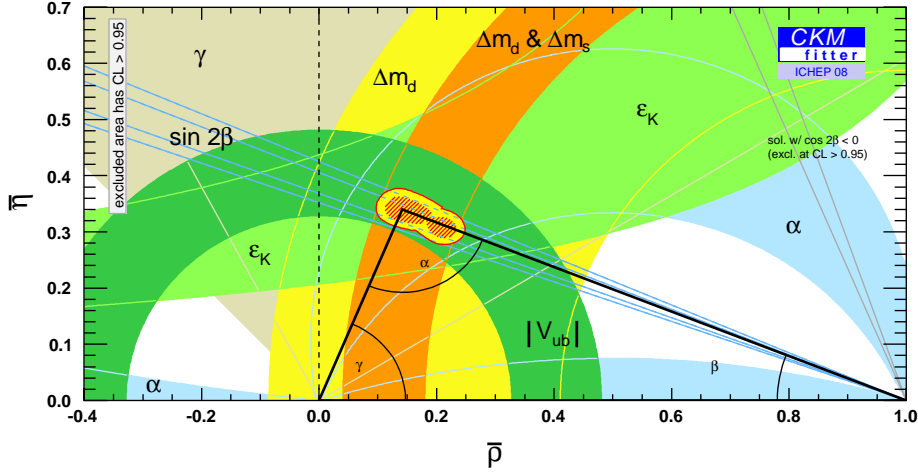}
\caption{Status of the unitarity triangle \cite{Charles:2004jd}.} 
\label{fig:ut}
\end{figure*}

Of central importance for flavor physics is the determination of
the CKM unitarity triangle. The current status is shown in Fig.~\ref{fig:ut}. 
Different determinations of the CKM parameters test the consistency 
of the SM and their precise values are needed as input for rare decays
with sensitivity to new physics.
Theoretical studies are crucial to translate measurements into
constraints on the unitarity triangle. An exceptionally clean quantity
is $\sin 2\beta$, determined from CP violation in $B\to\psi K_S$ with
negligible hadronic uncertainties. On the other hand, constraints from
$B$-$\bar B$ mixing ($\Delta m_{s,d}$) require input from lattice QCD and
the extraction of $\alpha$ and $\gamma$ benefits from progress in the
theory of hadronic $B$-decays. We will focus on the latter topic, which
has a wide range of applications both in two-body hadronic as well as in
exclusive rare and radiative decays of $B$ mesons.     
The remainder of this talk is organized into the sections: theory
of hadronic $B$ decays, phenomenology of $B\to M_1M_2$, 
next-to-next-to-leading order (NNLO) calculations
in QCD factorization, further highlights, and conclusions.

A comprehensive discussion of $B$, $D$, and $K$ decays in the era of
the LHC, covering both theoretical and experimental aspects,
can be found in the recent review \cite{Buchalla:2008jp}. 
We refer to this article for further details on the present discussion,
for an account of new physics scenarios and their implications
and for an extensive list of references.  
Prospects for $B$ physics at the LHC are also treated
in \cite{Fleischer:2008uj}.

\section{THEORY OF HADRONIC B DECAYS}
\label{sec:bdectheory}

\subsection{General Framework}

To be specific we will discuss two-body hadronic $B$ decays, but the
same theoretical methods also apply to rare and radiative processes such as
$B\to K^*\mu^+\mu^-$ or $B\to\rho\gamma$.

The computation of the amplitude for the decay of a $B$ meson into a
pair of light, charmless mesons, $B\to M_1M_2$, starts from the effective
weak Hamiltonian at the scale $\mu\approx m_b$ \cite{Buchalla:1995vs}
\begin{equation}\label{heff}
{\cal H}^{\Delta S=0}_{eff}=\frac{G_F}{\sqrt{2}}\left[
\lambda_u\left(\sum_{j=1,2}C_j Q^u_j+\sum_{penguins}C_P Q_P\right)+
\lambda_c\left(\sum_{j=1,2}C_j Q^c_j+\sum_{penguins}C_P Q_P\right)\right]
\end{equation}
$\lambda_p=V_{pb}V^*_{pd}$ ($p=u,c$) are CKM factors.
The Hamiltonian for $\Delta S=1$ transitions is obtained from the
$\Delta S=0$ case by interchanging $d\leftrightarrow s$.
The $C_i$ are Wilson coefficients, which include the physics at large scales,
$M_W$, $m_t$, or some new physics scale $M_{NP}$, down to the scale
$\mu\approx m_b$. They can be reliably computed in renormalization
group (RG) improved perturbation theory and are routinely used at
next-to-leading order (NLO).
The $Q_i$ are local operators of dimension 6, for instance
$Q^p_1=(\bar pb)_{V-A}(\bar dp)_{V-A}$.
The evaluation of their matrix elements
$\langle M_1M_2|Q_i|\bar B\rangle$ is the essential problem in the
theory of hadronic $B$ decays. The matrix elements contain three relevant
scales: $m_b$, $\sqrt{\Lambda_{QCD}m_b}$, $\Lambda_{QCD}$. Because
of the large hierarchy $m_b\gg\Lambda_{QCD}$ a systematic factorization
of the corresponding contribution is possible, and the hard ($\sim m_b$)
and hard-collinear ($\sim\sqrt{\Lambda m_b}$) parts may be treated in
perturbation theory. The nonperturbative effects from scales 
$\sim\Lambda_{QCD}$ can thus be separated from the perturbative ones and
the computation simplifies considerably.
While the effective Hamiltonian in (\ref{heff}) achieves a factorization
of high-energy scales $\sim M_W$ (Wilson coefficients) from the $m_b$
scale contained in the matrix elements $\langle Q_i\rangle$, the theory of
the matrix elements in the heavy-quark limit extends the factorization of
scales down to $\Lambda_{QCD}$.

The effective Hamiltonian, together with the matrix elements of the operators
$Q_i$, dictates the structure of the various $B$-decay amplitudes.
For instance, in the important case of $B$ decays into a pair of
light (charmless) mesons $M_1M_2$ one has schematically
\begin{equation}\label{amptp}
A(\bar B\to M_1M_2)=\lambda_u T + \lambda_c P
\end{equation}
This defines the tree and penguin amplitudes $T$ and $P$, respectively.
Their interplay, in conjunction with the magnitude and weak phase of the
CKM factors determines the phenomnology of all charmless $\Delta S=0$ and
$\Delta S=1$ transitions.

\subsection{Factorization of Matrix Elements for $B\to M_1M_2$}

\begin{figure*}[t]
\centering
\includegraphics[width=135mm]{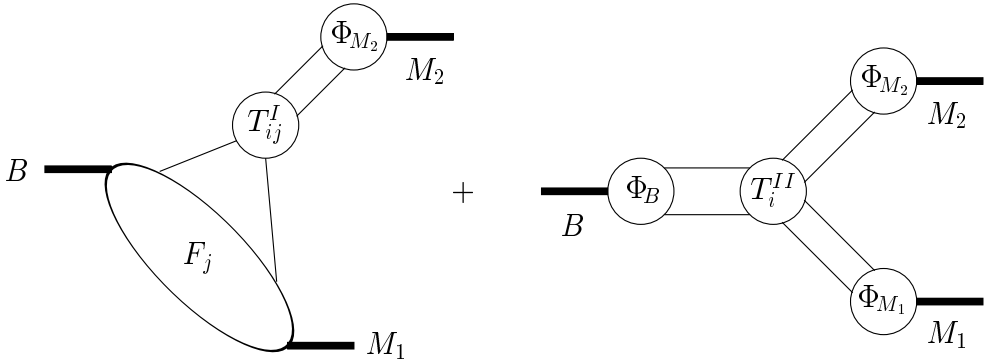}
\caption{Factorization formula.}
\label{fig:fform}
\end{figure*}

The factorization formula \cite{Beneke:1999br,Beneke:2000ry} 
for matrix elements $\langle M_1M_2|Q_i|\bar B\rangle$ 
is shown graphically in Fig.~\ref{fig:fform}.
It holds to leading order in $\Lambda_{QCD}/m_b$ and to all orders
in $\alpha_s$. The kernels $T^I$, $T^{II}$ are determined by the hard
scale $m_b$ ($T^I$) and the hard-collinear scale
$\sqrt{m_b\Lambda_{QCD}}$ ($T^{II}$). They are therefore calculable
in perturbation theory. $T^{II}$ describes the hard spectator interactions
and starts at order $\alpha_s$, whereas $T^I$ starts at order unity.
The kernels $T^{I,II}$ are systematically separated (factorized) from the
nonperturbative form factors $F_j$ and light-cone distribution amplitudes
$\Phi_B$, $\Phi_M$. These hadronic quantities are universal properties
of the mesons, or the $B\to M$ transition. The factorization formula thus
achieves a substantial simplification of the $B\to M_1M_2$ matrix elements,
which can be systematically improved in perturbation theory.
It is the basis for the theory of two-body hadronic $B$ decays.

The factorization theorem can be formulated using soft-collinear
effective theory (SCET) \cite{Bauer:2000yr,Bauer:2001yt}. 
This formalism is useful for proving factorization
and for disentangling the hard and hard-collinear scale in explicit terms. 
QCD factorization and SCET are theoretical concepts that are fully
consistent with each other, but they pertain to somewhat different aspects
of the problem of $B$-decay matrix elements. QCD factorization 
\cite{Beneke:1999br,Beneke:2000ry} refers to
the separation of the matrix elements into simpler long-distance
quantities and calculable hard interactions, while SCET is a general
effective field theory formulation for the relevant QCD modes
(hard, hard-collinear, collinear, soft) in the problem.
A somewhat analogous relation exists between the heavy-quark expansion
(HQE) and heavy-quark effective theory (HQET) in their application
to {\it inclusive\/} $B$ decays.
It should be emphasized that in fact there is a single common theory of
hadronic $B$ decays, which is based on factorization in the heavy-quark limit.
This should not be confused with differences in the phenomenological 
implementation of the theory that exist in the literature.
Those concern, for example, the treatment of power corrections or the use 
of experimental input for certain quantities.
A further variant of the basic idea of factorization is the so-called
PQCD approach \cite{Keum:2000wi}. 
It is more ambitious in the attempt to calculate further
hadronic quantities such as form factors or annihilation graphs. This
comes, however, at the expense of additional assumptions
($k_T$ factorization, Sudakov factors), which have been
critizised in the literature \cite{DescotesGenon:2001hm}.

\begin{figure*}[ht]
\centering
\includegraphics[width=34mm]{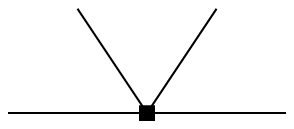}
\caption{Leading order diagram.}
\label{fig:lo}
\end{figure*}

\begin{figure*}[ht]
\centering
\includegraphics[width=135mm]{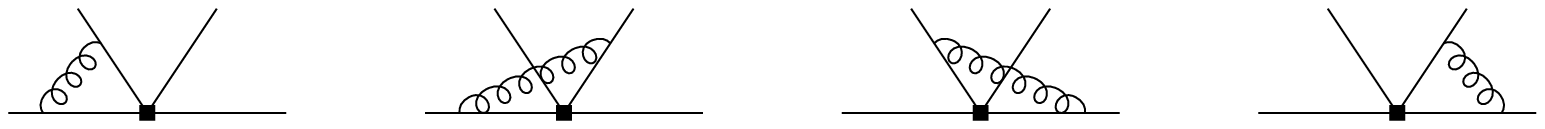}
\caption{${\cal O}(\alpha_s)$ vertex corrections.}
\label{fig:vertex}
\end{figure*}

\begin{figure*}[ht]
\centering
\includegraphics[width=68mm]{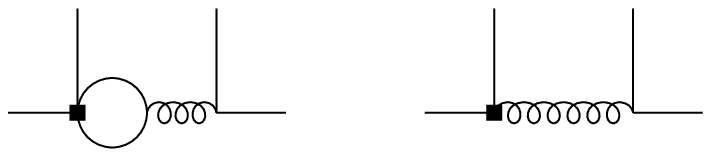}
\caption{${\cal O}(\alpha_s)$ penguin corrections.}
\label{fig:penguin}
\end{figure*}

\begin{figure*}[ht]
\centering
\includegraphics[width=68mm]{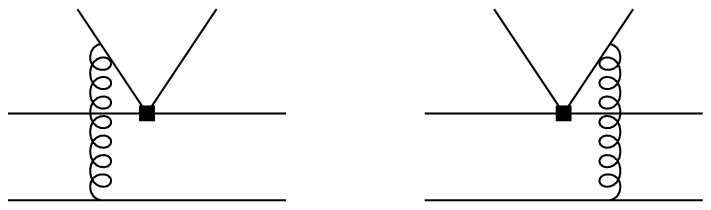}
\caption{${\cal O}(\alpha_s)$ hard spectator interactions.}
\label{fig:hardspec}
\end{figure*}

\begin{figure*}[ht]
\centering
\includegraphics[width=135mm]{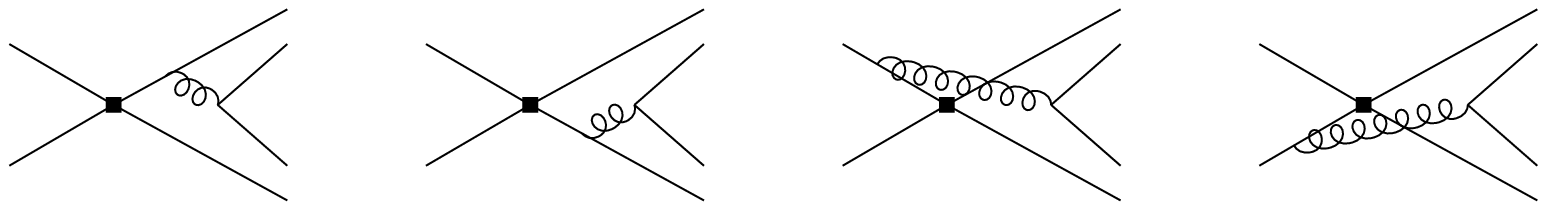}
\caption{${\cal O}(\Lambda/m_b)$ annihilation effects.}
\label{fig:annh}
\end{figure*}

A graphical representation of the various contributions to a
matrix element $\langle M_1M_2|Q_i|\bar B\rangle$ within
QCD factorization is given in Figs. \ref{fig:lo}-\ref{fig:hardspec}.
The leading order contribution is depicted in Fig. \ref{fig:lo},
vertex and penguin-type NLO corrections to the kernel $T^I$
are shown in Figs. \ref{fig:vertex} and \ref{fig:penguin},
respectively, and the NLO (${\cal O}(\alpha_s)$) contributions
to the hard-spectator scattering kernel $T^{II}$ are illustrated
in Fig. \ref{fig:hardspec}. 
Annihilation effects, shown in Fig. \ref{fig:annh},
have a power suppression $\sim\Lambda_{QCD}/m_b$ with respect to
the contributions listed before. They are not calculable within the
standard framework, but they may be estimated using model descriptions
and should be part of the analysis of theory uncertainties. 

\section{PHENOMENOLOGY OF $B\to M_1M_2$}
\label{sec:pheno}

\subsection{General Remarks}

All two-body $B$ decays into light pseudoscalar ($P$) and
longitudinal vector mesons ($V_L$), which are fully calculable
at leading power in QCD factorization, have been systematically
analyzed at NLO in $\alpha_s$.
The complete sets of decay modes
$B\to PP$ \cite{Beneke:2001ev,Beneke:2003zv}, 
$B\to PV$ \cite{Beneke:2003zv} (where $V$ is necessarily
longitudinally polarized) and $B\to V_LV_L$ \cite{Beneke:2006hg,BBK}
are thus available in the literature.
Transverse polarization observables for $B\to VV$ modes
are not entirely calculable in factorization. They have been discussed
in \cite{Beneke:2006hg,Kagan:2004uw}. Final states with vector and
axial vector mesons were considered recently in \cite{Cheng:2008gxa}.

The multitude of decay channels is a big asset for the phenomenology
of hadronic $B$ decays. First, symmetry relations
can be used to reduce the dependence on hadronic physics.
Second, different modes are in general sensitive to different
decay mechanisms and this may be exploited to selectively test
the theoretical understanding of QCD in these processes.     
In order to illustrate this we show in Table \ref{tab:ampl}
typical examples of decay modes that have a characteristic
dependence on specific decay topologies.
\begin{table}[hb]
\begin{center}
\caption{Amplitude topologies and flavor composition.}
\begin{tabular}{|l|c|c|}
\hline 
tree dominated & $\bar B_d\to\pi^+\pi^-$, $\rho^+\rho^-$ &
   $b(\bar d)\to d\bar uu(\bar d)$\\
\hline
pure tree & $B^-\to\pi^-\pi^0$, $\rho^-\rho^0$ &
   $b(\bar u)\to d\bar uu(\bar u)$ (isospin)\\
\hline
pure penguin & $B^-\to\bar K^0\pi^-$ &
   $b(\bar u)\to s\bar dd(\bar u)$\\
\hline
pure annihilation & $\bar B_d\to K^+K^-$ &
   $b(\bar d)\to u\bar ss \bar u$\\
\hline
\end{tabular}
\label{tab:ampl}
\end{center}
\end{table}
The first two lines show $B$ decays that can be generated by
tree diagrams. Here we distinguish the case where other, subdominant
contributions (e.g. penguins) are possible from the ``pure-tree''
case, where those are forbidden by isospin. (In the latter
case electroweak penguins, which violate isospin, can still appear.)
The possible topologies can be read off from the flavor
structure of the transition, which is given in the last column.
The flavor of the spectator quark is indicated in brackets.

From the comparison of theoretical results for $B\to M_1 M_2$ with
available data one finds a good overall agreement.
However, the general picture, or the consideration of global fits,
may conceal particular deviations, either due to hadronic
physics effects or from flavor dynamics beyond the SM.
It is therefore mandatory to investigate and assess specific
observables. We shall come back to this issue with the
discussion of an example in section \ref{subsec:pfp}.
In this context it is also important to aim, as far as possible,
at a separation of hadronic and flavor physics effects by studying 
suitable observables. 

Focussing on the QCD aspects of hadronic $B$ decays, factorization is able 
to account, qualitatively at least and often in quantitative detail,
for several remarkable features of these processes. Among them one may
note: a) the relative size of pure penguin and tree amplitudes;
b) the hierarchy between the penguin amplitudes for $V_LV_L$ and
$PP$ final states; c) the apparent smallness of pure annihilation modes;
d) the small to moderate size of direct CP asymmetries, which are
suppressed by strong phases predicted to be of order $\alpha_s$ or
$\Lambda/m_b$.

\begin{figure*}[ht]
\centering
\includegraphics[width=100mm]{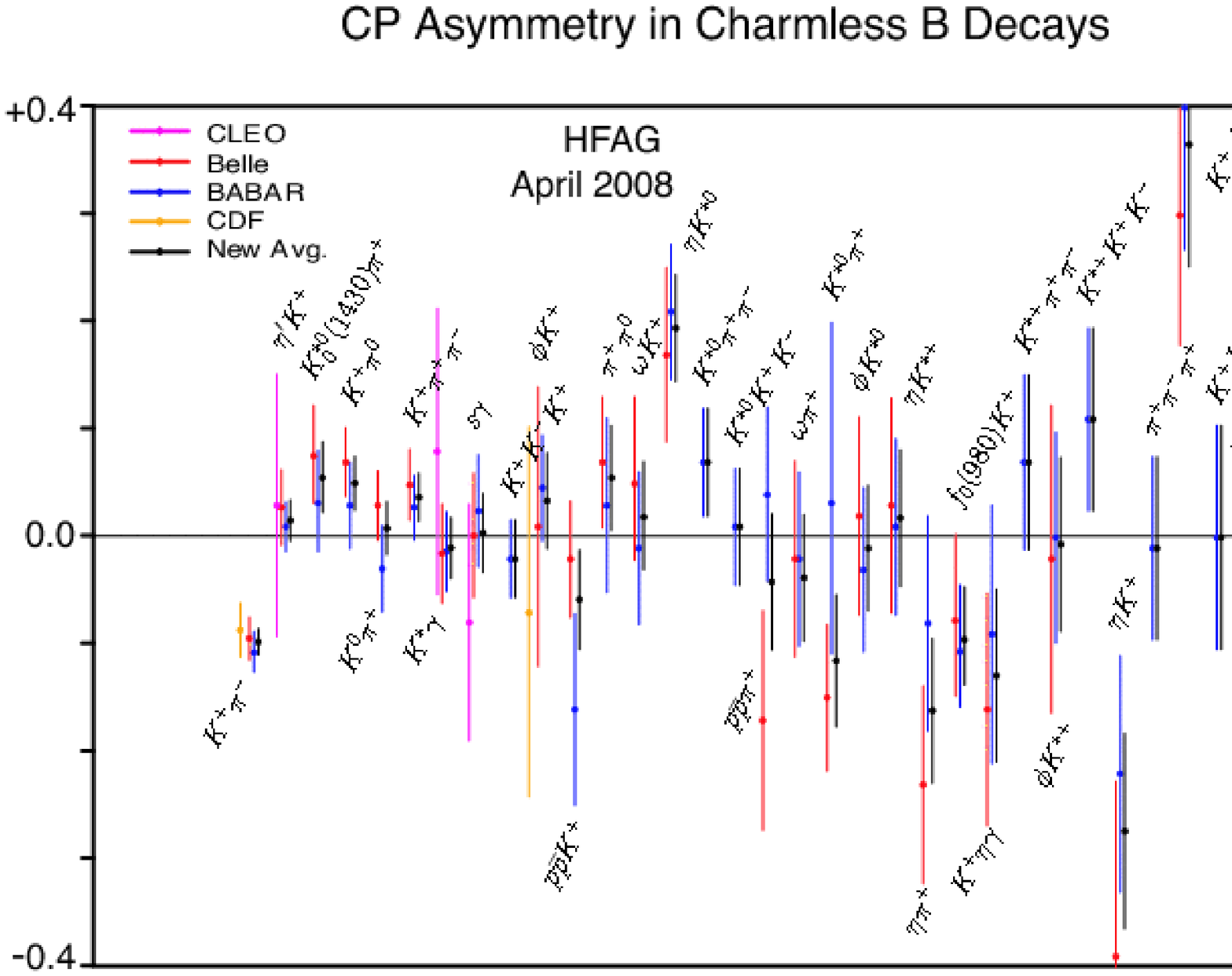}
\caption{Direct CP asymmetries in hadronic $B$ decays \cite{Barberio:2008fa}.}
\label{fig:acpdir}
\end{figure*}

The last point d) is rather qualitative in nature. Since power
corrections are likely to compete with the calculable ${\cal O}(\alpha_s)$
effects, the strong phases and hence direct CP asymmetries are very
difficult to predict with good precision. Nevertheless, some trace
of the expected parametric suppression should be visible in the
data. In Fig. \ref{fig:acpdir} we reproduce a compilation from HFAG
\cite{Barberio:2008fa} of the
most precisely measured direct CP asymmetries. 
Most of the asymmetries are below $10\%$ in magnitude, and all
are compatible with being smaller than $20\%$.
This quick glance can clearly be only a rough indication.
One has to note for instance that some asymmetries are expected
to (almost) vanish because of the (near) absence of a second, 
interfering amplitude. Examples are $B^+\to\pi^+\pi^0$ or $K^0\pi^+$. 
Also, experimental errors are still large in many cases. 
The most precisely measured direct CP asymmetry is
\begin{equation}\label{acpkppm}
A_{CP}(K^+\pi^-)=-0.097\pm 0.012
\end{equation}
It is interesting to estimate the strong phase difference implied by
this result (assuming the SM).
We may write
\begin{equation}\label{acpth}
A_{CP}(K^+\pi^-)\equiv
\frac{\Gamma(\bar B_d\to K^-\pi^+)-\Gamma(B_d\to K^+\pi^-)}{
\Gamma(\bar B_d\to K^-\pi^+)+\Gamma(B_d\to K^+\pi^-)}
=\frac{2d\sin\gamma\sin\phi}{1+d^2-2d\cos\gamma\cos\phi}
\approx 2d\sin\gamma\sin\phi
\end{equation}
where $\gamma$ is the CKM angle, $\phi$ is the strong phase
difference between tree and penguin amplitude and 
\begin{equation}\label{ddef}
d\equiv\left|\frac{T}{P}\right|\, 
\left|\frac{V_{ub} V_{us}}{V_{cb} V_{cs}}\right|
\end{equation}
with $T$ and $P$ the hadronic tree and penguin amplitudes as defined
in (\ref{amptp}). The approximation in the last relation of (\ref{acpth})
holds to first order in the small quantity $d$.
Using $|P/T|=0.092$ from QCD factorization \cite{Beneke:2001ev} and 
$|(V_{ub} V_{us})/(V_{cb} V_{cs})|=0.021$, we have $d=0.23$.
For a typical value of $\gamma=67^\circ$, $\sin\gamma=0.92$, we then
find from (\ref{acpth}) for the strong phase difference $\phi$ 
in $B\to K^+\pi^-$
\begin{equation}\label{phikpi}
\sin\phi=-0.23 \qquad\quad \phi=-13^\circ
\end{equation}
The sign of $\phi$ comes out opposite to the expectation from
perturbation theory at order $\alpha_s$.
On the other hand the strong phase difference is rather moderate
and compatible with the expected parametric suppression.

The direct CP asymmetry in $B\to K^+\pi^0$ is also small, but it appears
to have the opposite sign compared to (\ref{acpkppm}). 
HFAG \cite{Barberio:2008fa} quotes $A_{CP}(K^+\pi^0)=0.050\pm 0.025$.
The origin of the difference to (\ref{acpkppm}) is not entirely clear 
at present.
It could be due to QCD effects or, in principle, also be a signal of
new physics. The interest in this issue has been stressed  
in \cite{:2008zza}. Because it is difficult to obtain
accurate theoretical results for direct CP asymmetries, the resolution
of this question appears challenging.

Among other subjects that require further attention is
the direct CP asymmetry in $B\to\pi^+\pi^-$. Here the result of Belle
could point to substantial strong phases, while the measurement 
by BaBar is more in line with theory expectations. For more details, see
also the discussion in sec. \ref{subsec:pfp}.
 
We finally remark that the measured branching ratio of $B\to\pi^0\pi^0$
is somewhat high compared to typical estimates in QCD factorization,
which however have large uncertainties for this channel 
(see sec. \ref{sec:nnlo}).
The agreement is better in the similar case of $B\to\rho^0_L\rho^0_L$.

\subsection{Precision Flavor Physics}
\label{subsec:pfp}

The most promising applications of the theory of hadronic
$B$ decays are those where the dependence on QCD dynamics is not the
dominant feature. More suitable in that respect than, for instance,
direct CP asymmetries or absolute branching fractions are
flavor observables with small dependence on hadronic physics,
often involving only ratios of amplitudes.
In such cases even a moderate, but realistic level of accuracy in the
hadronic dynamics can lead to high precision in the determination
of flavour physics parameters.
In order to illustrate this point we discuss one particular example,
the extraction of CKM parameters from CP violation in $B\to\pi^+\pi^-$
or $B\to\rho^+_L\rho^-_L$ decays.

The time dependent CP asymmetry in $B\to\pi^+\pi^-$ reads
\begin{equation}\label{acpsc}
a_{CP}(t)=\frac{\Gamma(\bar B(t)\to\pi^+\pi^-)-\Gamma(B(t)\to\pi^+\pi^-)}{
\Gamma(\bar B(t)\to\pi^+\pi^-) + \Gamma(B(t)\to\pi^+\pi^-)}=
S\, \sin(\Delta M t) -C\, \cos(\Delta M t)
\end{equation}
The form of the decay amplitude is given by eq. (\ref{amptp}).
The $S$ term in (\ref{acpsc}) arises from the interference
of $B$-$\bar B$ mixing with the dominant tree amplitude $\sim T$,
with a small correction due to the penguin component $\sim P$.
The $C$ term signals direct CP violation and comes from the
tree-penguin interference in (\ref{amptp}). The formulas for 
$B\to\pi^+\pi^-$ apply also to the case of $B\to\rho^+_L\rho^-_L$.
The measured CP violation parameters for both channels
are given in Table \ref{tab:cpvsc}.
\begin{table}[hb]
\begin{center}
\caption{CP violation in $B\to\pi^+\pi^-$, $\rho^+_L\rho^-_L$
\cite{Barberio:2008fa}.}
\begin{tabular}{|c|c|c|}
\hline
 & $S$ & $C$\\
\hline
$B\to\pi^+\pi^-$ & $-0.61\pm 0.08$ & $-0.38\pm 0.07^*$\\
\hline
$B\to\rho^+_L\rho^-_L$ & $-0.05\pm 0.17$ & $-0.06\pm 0.13$\\
\hline
\end{tabular}
\label{tab:cpvsc}
\end{center}
\end{table}
For $C_\pi$ the measurements of BaBar and Belle are not in very
good agreement. BaBar finds $C_\pi=-0.21\pm 0.09$, Belle
obtains $C_\pi=-0.55\pm 0.09$. Table \ref{tab:cpvsc} quotes
the average without inflating the error.
The Belle result appears somewhat large with respect to theory
expectations, whereas the BaBar number is perfectly compatible.
Independent measurements at LHCb will be very important.

Up to an irrelevant constant factor the amplitude
in (\ref{amptp}) can, for $\bar B\to\pi^+\pi^-$, also be written as
\begin{equation}\label{amppppm}
A(\bar B\to\pi^+\pi^-)\sim \sqrt{\bar\rho^2+\bar\eta^2} \, e^{-i\gamma}
+ r_\pi e^{i\phi_\pi}
\end{equation}
which defines magnitude $r=r_\pi$ and phase $\phi=\phi_\pi$ of the
penguin-to-tree ratio. $\bar\rho$, $\bar\eta$ are 
Wolfenstein parameters and $\gamma$ is a CKM phase. 
A similar expression holds for $\bar B\to\rho^+_L\rho^-_L$
with the replacement of ($r_\pi$, $\phi_\pi$) by ($r_\rho$, $\phi_\rho$).

Knowledge of $S$ and  of the CKM angle $\beta$ or, equivalently,
\begin{equation}\label{taubeta}
\tau\equiv\cot\beta=2.54\pm 0.13
\end{equation}
from CP violation in $B\to\psi K_S$ \cite{Barberio:2008fa}, fixes the 
unitarity triangle in the form \cite{Buchalla:2003jr,Buchalla:2004tw}
\begin{equation}\label{rhotau}
\bar\rho=1-\tau\bar\eta
\end{equation}
\begin{equation}\label{etataus}
\bar\eta=\frac{1+\tau S-\sqrt{1-S^2}}{(1+\tau^2)S}
\,\left(1 + r\cos\phi \right)
\end{equation}
Up to the hadronic correction $r\cos\phi$ from the penguin-to-tree ratio
the unitarity triangle is entirely determined by the observables
$S$ and $\tau=\cot\beta$.
From QCD factorization we have \cite{BBK,Buchalla:2004tw}
\begin{eqnarray}\label{rpirrho}
r_\pi &=& 0.11\pm 0.03\\
r_\rho &=& 0.038\pm 0.024
\end{eqnarray}

The analysis can be performed in the same way for the $\pi^+\pi^-$ 
or the $\rho^+_L\rho^-_L$ case.
The results are compatible within errors. Here we will concentrate on 
$B\to\rho^+_L\rho^-_L$, which from a theoretical point of view appears 
slightly prefered because of a smaller size of the penguin amplitude and 
because all three $B\to\rho_L\rho_L$ decays are very well described
within QCD factorization \cite{Beneke:2006hg,BBK}.
Also, for direct CP violation in $B\to\pi^+\pi^-$ the experimental
situation is still not completely clarified, as mentioned above,
whereas $C_\rho$ is measured to be small in agreement with theory
expectations. 

For $B\to\rho^+_L\rho^-_L$ we have $\cos\phi_\rho\approx 1$, which
corresponds to a small phase, as confirmed empirically
by the small direct CP violation $C_\rho$.
From the parameters $\tau$, $S_\rho$, $r_\rho$ specified above we then 
obtain $\bar\rho$ (\ref{rhotau}) and $\bar\eta$ (\ref{etataus}).
This implies a value for $\gamma=\arctan(\bar\eta/\bar\rho)$,
\begin{equation}\label{gammares}
\gamma=72.4^\circ\pm 1.3^\circ\, (\tau)\, \pm 5.1^\circ\, (S_\rho)\,
\pm 3.2^\circ\, (r_\rho)
\end{equation}
The uncertainty is still domianted by the experimental error in $S_\rho$.
The theoretical quantity $r_\rho$ may be cross-checked with the help
of other measurements. The branching fraction for the pure penguin
decay $B^-\to\bar K^{*0}_L\rho^-_L$ \cite{Beneke:2006rb}
can be used to extract a penguin-to-tree ratio $r_{K^*\rho}\approx 0.04$ 
(defined similarly as $r_\rho$) in agreement with QCD factorization.
Employing $SU(3)$ symmetry, $r_\rho\approx 0.06$ can be obtained from
$\bar B_d\to\bar K^{*0}_LK^{*0}_L$ \cite{BBK}.
Improved measurements of $\bar B_d\to\bar K^{*0}_LK^{*0}_L$ could
help to further reduce the theory uncertainty in (\ref{gammares}).

\section{NNLO EFFECTS IN QCD FACTORIZATION}
\label{sec:nnlo}

For several important classes of contributions in charmless
two-body $B$ decays the perturbative calculations in QCD factorization
have recently been extended to the level of NNLO corrections.
By NNLO we here mean the effects of order $\alpha^2_s$.
In many cases this level of accuracy is probably below the
size of uncertainties from other sources, most notably from power
corrections. However, the explicit knowledge of NNLO corrections
is of some conceptual interest as it extends the factorization formula
to the next nontrivial level in perturbation theory.
In addition, there are quantities for which the NNLO effects are
likely to be also numerically important. These are cases where
a contribution is absent at ${\cal O}(1)$ and thus the 
${\cal O}(\alpha_s)$ term, a NLO contribution in the general
counting scheme, is effectively the lowest order. An example is
hard spectator scattering. Another case is the coefficient
of so-called color-suppressed amplitudes, $a_2$, which is
accidentally small at leading and next-to-leading order and therefore
rather sensitive to NNLO effects.
 
We briefly summarize the NNLO corrections that have been computed so far.
These are, first, the
${\cal O}(\alpha^2_s)$, one-loop hard spectator interactions
for current-current operators \cite{Beneke:2005vv,Kivel:2006xc,Pilipp:2007mg}
and for penguin contributions \cite{Beneke:2006mk}.
The calculations in \cite{Beneke:2005vv,Kivel:2006xc} are done in
the SCET framework, whereas \cite{Pilipp:2007mg} employs factorization
in full QCD. The agreement of the results is an example of the
equivalence of the two formulations.
Second, the two-loop vertex corrections ($T^I$) have been addressed
for the first time in \cite{Bell:2007tv}, where the imaginary
part is computed explicitly. The real part is still unpublished. 

The numerical impact of NNLO effects in comparison with other 
corrections for the coefficient $a_2(\pi\pi)$, which dominates the
amplitude  for $B\to\pi^0\pi^0$, is illustrated by the
following compilation \cite{Buchalla:2008jp}:
\begin{eqnarray}
a_2(\pi\pi) &=& 0.184 -[0.153+0.077 i]_{\alpha_s}+[?-0.049 i]_{\alpha^2_s}
\nonumber\\
&&+\frac{9 f_{M_1}f_B}{0.485\, m_b\,\lambda_B\, F^{B\to M_1}}
\left( [0.122]_{\alpha_s}+[0.050+0.053 i]_{\alpha^2_s}+[0.071]_{\rm twist-3}
\right) \nonumber\\
&=& 0.275^{+0.228}_{-0.135}+\left(-0.073^{+0.115}_{-0.082}\right)i
\label{a2nnlo}
\end{eqnarray}
The first line lists the vertex corrections at various orders.
The still unknown real part at order $\alpha^2_s$ is indicated
by the question mark. The second line shows the amplitude
from hard spectator interactions. It includes an estimate of
power corrections from twist-3 contributions to the 
light-cone wave function of the pion (last term).
The normalization of the prefactor is such that it gives unity
for default values of the parameters. The third line displays the
full result together with the total uncertainty.

In (\ref{a2nnlo}) the cancellation of leading and next-to-leading
vertex contributions is clearly visible. This implies the dominance of
the hard spectator term and the relativly large impact of NNLO
effects and power corrections. In addition, the important parameter
$\lambda_B$, which characterizes spectator scattering, is not
well known. An accurate prediction of the $B\to\pi^0\pi^0$ branching 
ratio is therefore difficult. One may note that the NNLO term in
the hard spectator interaction enhances the size of $a_2(\pi\pi)$,
bringing the theory prediction for $B(B\to\pi^0\pi^0)$ closer to
the relatively large experimental result.

\section{FURTHER HIGHLIGHTS}
\label{sec:highl}

We briefly comment on some examples of further topics, which
are of great interest for the Tevatron and the LHC. 
A comprehensive account is contained in \cite{Buchalla:2008jp}.

\subsection{$B_s\to\mu^+\mu^-$}

In the SM the branching fraction for the leptonic FCNC decay
$B_s\to\mu^+\mu^-$ can be represented as
\begin{equation}\label{bsmumu}
B(B_s\to\mu^+\mu^-)=3.9\cdot 10^{-9}\, 
\left[\frac{f_{B_s}}{240\,{\rm MeV}}\right]^2
\end{equation}
The branching fraction is sensitive to the decay constant
$f_{B_s}$. Apart from that it is completely dominated by
short distance physics from top-quark loops and can be predicted
with very good accuracy. In the SM the branching ratio is
very small because of a helicity suppression. This leaves room
for large enhancements from new physics, which are possible for instance  
in the Minimal Supersymmetric Standard Model with large $\tan\beta$,
and could saturate the experimental bound. 
The current upper limit from CDF is \cite{Aaltonen:2007kv}
\begin{equation}\label{bsmumuexp}
B(B_s\to\mu^+\mu^-) < 47\cdot 10^{-9}
\qquad {\rm at} \quad 90\% \, \, {\rm C.L.}
\end{equation}

Since $B_s\to\mu^+\mu^-$ is theoretically very clean, 
has interesting sensitivity
to new physics and a clear experimental signature, it is a prime
goal for flavor studies at hadron colliders.

\subsection{$B\to K^*l^+l^-$}

Rich possibilities to study short-distance flavor physics are
also associated with the semileptonic rare decay $B\to K^*l^+l^-$.
Besides the differential branching ratio various other kinematical 
distributions may be measured in this decay. This would help to disentangle
different sources of new physics. An interesting observable is,
for example, the position of a zero in the forward-backward
asymmetry $A_{FB}$ as a function of the dilepton mass $q^2$.
$A_{FB}$ is defined as the rate asymmetry between forward and backward
going leptons $l^+$ in the dilepton center-of-mass frame.
The position $q^2=q^2_0$ of the zero in $A_{FB}(q^2)$ has the form 
\begin{equation}\label{afbzero}
q^2_0=\frac{F^{B\to K^*}_1}{F^{B\to K^*}_2} \frac{C_{7\gamma}}{C^{\rm eff}_9}
\end{equation}  
In the heavy-quark limit a SCET relation ensures that unknown hadronic
physics cancels in the ratio of form factors $F^{B\to K^*}_{1,2}$. 
This leads to a prediction
of $q^2_0$ only in terms of calculable short-distance quantities 
($\sim C_{7\gamma}/C^{\rm eff}_9$).
NLO QCD corrections have been computed in \cite{Beneke:2001at}.

\subsection{$B_s$-$\bar B_s$ Mixing, $B_s\to\psi\phi$, $B_s\to\phi\phi$}

Mixing and CP violation in the $B_s$-$\bar B_s$ system can be
investigated using decay asymmetries and distributions of the
process $B_s\to\psi\phi$. In some sense this decay is the analogue
of $B_d\to\psi K_S$ in the $B_d$ system, but the vector-vector
final state and the larger width difference $\Delta\Gamma_s$ in the
$B_s$ case make the analysis of $B_s\to\psi\phi$ more involved
\cite{Dunietz:2000cr}.
An analysis of recent measurements by CDF and D0 has been
given in \cite{Bona:2008jn}, suggesting indications for discrepancies
with the SM. Definite conclusions about the presence
of new physics appear still premature. A further recent discussion
of related topics can be found in \cite{Gronau:2008hb}.

\section{CONCLUSIONS}
\label{sec:concl}

Flavor studies with $K$, $D$ and, in particular, $B$-meson
decays complement direct searches for new physics at colliders.
Hadron machines deliver very large samples of $B$ mesons.
Their decay modes can be extracted from background if the final state is 
sufficiently clean, with charged tracks, typically few particles and one photon
at most. In addition to improved measurements of known $B_{u,d}$
decay channels, the investigation of the rich and largely unexplored field 
of $B_s$ decays is of special importance. 
Among the processes of interest are  
$B_s$-$\bar B_s$ mixing, $B_s\to\mu^+\mu^-$,
the exclusive rare decays $B\to K^*l^+l^-$, $B\to \rho\gamma$,
$B\to K^*\gamma$, $B_s\to\phi\gamma$, and hadronic two-body modes
such as $B\to\pi^+\pi^-$, $B_s\to K^+K^-$ or $B_s\to\phi\phi$.
For the theoretical treatment of these decays
and the control of hadronic physics it is necessary to use the systematic
methods of factorization, SCET, flavor symmetries, lattice QCD and
light-cone sum rules, which are in many ways complementary to each other.
Presumably the most reliable results will be obtained when theory 
calculations enter only in small corrections to clean flavor
observables. In these cases moderate precision in hadronic quantities
will suffice to reach high accuracy in quantities of flavor physics.
The large variety of $B$ decay channels will allow us to identify
observables with this property or to construct them by a combination
of several modes.
With the available theoretical tools many opportunities exist at hadron 
colliders for precision tests and discoveries in $B$ physics.

\begin{acknowledgments}
I thank the organizers for the invitation to this interesting
and enjoyable conference. This work (LMU-ASC 47/08) is supported in part 
by the DFG cluster of excellence ``Origin and Structure of the Universe''.
\end{acknowledgments}



\begin{thebibliography}{99} 


\bibitem{Kronfeld:2008zz}
  A.~S.~Kronfeld  [USQCD Collaboration],
  arXiv:0807.2220 [physics.comp-ph];
  E.~Gamiz,
  arXiv:0807.0381 [hep-ph].


\bibitem{Ball:2004ye}
  P.~Ball and R.~Zwicky,
  Phys.\ Rev.\  D {\bf 71} (2005) 014015
  [arXiv:hep-ph/0406232].

\bibitem{Ball:2004rg}
  P.~Ball and R.~Zwicky,
  Phys.\ Rev.\  D {\bf 71} (2005) 014029
  [arXiv:hep-ph/0412079].


\bibitem{Charles:2004jd}
  J.~Charles {\it et al.}  [CKMfitter Group],
  Eur.\ Phys.\ J.\  C {\bf 41} (2005) 1
  [arXiv:hep-ph/0406184];
  updates at: http://ckmfitter.in2p3.fr 

\bibitem{Buchalla:2008jp}
  M.~Artuso {\it et al.},
  arXiv:0801.1833 [hep-ph].

\bibitem{Fleischer:2008uj}
  R.~Fleischer,
  arXiv:0802.2882 [hep-ph].




\bibitem{Buchalla:1995vs}
  G.~Buchalla, A.~J.~Buras and M.~E.~Lautenbacher,
  Rev.\ Mod.\ Phys.\  {\bf 68} (1996) 1125
  [arXiv:hep-ph/9512380].





\bibitem{Beneke:1999br}
  M.~Beneke, G.~Buchalla, M.~Neubert and C.~T.~Sachrajda,
  Phys.\ Rev.\ Lett.\  {\bf 83} (1999) 1914
  [arXiv:hep-ph/9905312].

\bibitem{Beneke:2000ry}
  M.~Beneke, G.~Buchalla, M.~Neubert and C.~T.~Sachrajda,
  Nucl.\ Phys.\  B {\bf 591} (2000) 313
  [arXiv:hep-ph/0006124].

\bibitem{Bauer:2000yr}
  C.~W.~Bauer, S.~Fleming, D.~Pirjol and I.~W.~Stewart,
  Phys.\ Rev.\  D {\bf 63} (2001) 114020
  [arXiv:hep-ph/0011336].

\bibitem{Bauer:2001yt}
  C.~W.~Bauer, D.~Pirjol and I.~W.~Stewart,
  Phys.\ Rev.\  D {\bf 65} (2002) 054022
  [arXiv:hep-ph/0109045].

\bibitem{Keum:2000wi}
  Y.~Y.~Keum, H.~N.~Li and A.~I.~Sanda,
  Phys.\ Rev.\  D {\bf 63} (2001) 054008
  [arXiv:hep-ph/0004173].

\bibitem{DescotesGenon:2001hm}
  S.~Descotes-Genon and C.~T.~Sachrajda,
  Nucl.\ Phys.\  B {\bf 625} (2002) 239
  [arXiv:hep-ph/0109260].






\bibitem{Beneke:2001ev}
  M.~Beneke, G.~Buchalla, M.~Neubert and C.~T.~Sachrajda,
  Nucl.\ Phys.\  B {\bf 606} (2001) 245
  [arXiv:hep-ph/0104110].

\bibitem{Beneke:2003zv}
  M.~Beneke and M.~Neubert,
  Nucl.\ Phys.\  B {\bf 675} (2003) 333
  [arXiv:hep-ph/0308039].

\bibitem{Beneke:2006hg}
  M.~Beneke, J.~Rohrer and D.~Yang,
  Nucl.\ Phys.\  B {\bf 774} (2007) 64
  [arXiv:hep-ph/0612290].

\bibitem{BBK}
  M.~Bartsch, G.~Buchalla and C.~Kraus, LMU-ASC 43/08,
  in preparation.

\bibitem{Kagan:2004uw}
  A.~L.~Kagan,
  Phys.\ Lett.\  B {\bf 601} (2004) 151
  [arXiv:hep-ph/0405134].

\bibitem{Cheng:2008gxa}
  H.~Y.~Cheng and K.~C.~Yang,
  arXiv:0805.0329 [hep-ph].



\bibitem{Barberio:2008fa}
  E.~Barberio {\it et al.},
  arXiv:0808.1297 [hep-ex];
  updates at: http://www.slac.stanford.edu/xorg/hfag/


\bibitem{:2008zza}
    [The Belle Collaboration],
  Nature {\bf 452} (2008) 332.




\bibitem{Buchalla:2003jr}
  G.~Buchalla and A.~S.~Safir,
  Phys.\ Rev.\ Lett.\  {\bf 93} (2004) 021801
  [arXiv:hep-ph/0310218].

\bibitem{Buchalla:2004tw}
  G.~Buchalla and A.~S.~Safir,
  Eur.\ Phys.\ J.\  C {\bf 45} (2006) 109
  [arXiv:hep-ph/0406016].

\bibitem{Beneke:2006rb}
  M.~Beneke, M.~Gronau, J.~Rohrer and M.~Spranger,
  Phys.\ Lett.\  B {\bf 638} (2006) 68
  [arXiv:hep-ph/0604005].







\bibitem{Beneke:2005vv}
  M.~Beneke and S.~J\"ager,
  Nucl.\ Phys.\  B {\bf 751} (2006) 160
  [arXiv:hep-ph/0512351].

\bibitem{Kivel:2006xc}
  N.~Kivel,
  JHEP {\bf 0705} (2007) 019
  [arXiv:hep-ph/0608291].

\bibitem{Pilipp:2007mg}
  V.~Pilipp,
  Nucl.\ Phys.\  B {\bf 794} (2008) 154
  [arXiv:0709.3214 [hep-ph]].

\bibitem{Beneke:2006mk}
  M.~Beneke and S.~J\"ager,
  Nucl.\ Phys.\  B {\bf 768} (2007) 51
  [arXiv:hep-ph/0610322].

\bibitem{Bell:2007tv}
  G.~Bell,
  Nucl.\ Phys.\  B {\bf 795} (2008) 1
  [arXiv:0705.3127 [hep-ph]].





\bibitem{Aaltonen:2007kv}
  T.~Aaltonen {\it et al.}  [CDF Collaboration],
  Phys.\ Rev.\ Lett.\  {\bf 100} (2008) 101802
  [arXiv:0712.1708 [hep-ex]].


\bibitem{Beneke:2001at}
  M.~Beneke, T.~Feldmann and D.~Seidel,
  Nucl.\ Phys.\  B {\bf 612} (2001) 25
  [arXiv:hep-ph/0106067].

\bibitem{Dunietz:2000cr}
  I.~Dunietz, R.~Fleischer and U.~Nierste,
  Phys.\ Rev.\  D {\bf 63} (2001) 114015
  [arXiv:hep-ph/0012219].

\bibitem{Bona:2008jn}
  M.~Bona {\it et al.}  [UTfit Collaboration],
  arXiv:0803.0659 [hep-ph].

\bibitem{Gronau:2008hb}
  M.~Gronau and J.~L.~Rosner,
  arXiv:0808.3761 [hep-ph].




\end{thebibliography}
\end{document}